\documentclass[11pt,a4paper]{article}
\usepackage{amsmath,amssymb}

\usepackage{amsmath}

\usepackage{color}
\usepackage[colorlinks,linkcolor=blue,citecolor=red]{hyperref}

\DeclareMathSymbol{\bbbr}{\mathalpha}{AMSb}{"52}
\DeclareMathSymbol{\bbbc}{\mathalpha}{AMSb}{"52}

\newcommand{\cE}{\mathcal E}
\newcommand{\cM}{\mathcal M}

\newtheorem{theorem}{Theorem}

\begin{document}

\title{On a class of integrable systems of Monge-Amp\`ere type}

\author{B. Doubrov$^1$, E.V. Ferapontov$^2$, B. Kruglikov$^{3,4}$, V.S.  Novikov$^2$}
     \date{}
     \maketitle
     \vspace{-5mm}
\begin{center}
$^1$Department of Mathematical Physics\\
Faculty of Applied Mathematics\\
Belarussian State University\\
Nezavisimosti av.\ 4, 220030 Minsk, Belarus\\
  \ \\
$^2$Department of Mathematical Sciences \\ Loughborough University \\
Loughborough, Leicestershire LE11 3TU \\ United Kingdom \\
\ \\
$^3$Department of Mathematics and Statistics \\
UiT the Arctic University of Norway\\
Troms\o\ 90-37, Norway\\
\ \\
$^4$Department of Mathematics and Natural Sciences\\
University of Stavanger,
40-36 Stavanger, Norway\\
\ \\
e-mails: \\[1ex]  \texttt{doubrov@islc.org} \\
\texttt{E.V.Ferapontov@lboro.ac.uk}\\
\texttt{boris.kruglikov@uit.no} \\
\texttt{V.Novikov@lboro.ac.uk}
\end{center}

\vspace{1cm}

\begin{abstract}

We investigate a class of multi-dimensional two-component systems of Monge-Amp\`ere type that can be viewed as  generalisations of heavenly-type equations appearing in  self-dual Ricci-flat geometry. Based on the  Jordan-Kronecker theory of skew-symmetric matrix pencils, a classification of normal forms of such systems is obtained.
All two-component systems of Monge-Amp\`ere type turn out to be integrable, and can be represented as the commutativity conditions of  parameter-dependent vector fields. 

Geometrically, systems of Monge-Amp\`ere type are associated with linear sections of the Grassmannians. This leads to 
an invariant differential-geometric characterisation of the Monge-Amp\`ere property.

\bigskip


\bigskip

\end{abstract}

\newpage


\newpage

\section{Introduction}

Let  $u({\bf x})$ and $ v({\bf x})$ be functions of $d$ independent variables ${\bf x}=(x^1, \dots, x^d)$. In paper \cite{DFKN} we have initiated the study of integrability of first-order  systems of the form
\begin{equation}
 F(u_1, \dots, u_d, v_1, \dots, v_d)=0, ~~~ G(u_1, \dots, u_d, v_1, \dots, v_d)=0,
\label{d}
\end{equation}
where $F, G$ are (nonlinear) functions of the  partial derivatives $u_i=\frac{\partial u}{\partial x^i}, \ v_i=\frac{\partial v}{\partial x^i}$. The geometry behind systems (\ref{d}) is as follows. Let $V$ be the $(d+2)$-dimensional vector space with coordinates $ x^1,  \dots,  x^d,  u,  v$. Solutions to system (\ref{d}) correspond to $d$-dimensional submanifolds of $V$ defined as  $u=u({\bf x}), \ v= v({\bf x})$. Their $d$-dimensional tangent spaces, specified by the equations $du=u_idx^i, \ dv=v_idx^i$, are parametrised by $2\times d$ matrices 
$$
U=\left( \begin{array}{ccc}
 u_1 & \dots &u_d
 \\
v_1 & \dots  &v_d
 \end{array}\right),
$$
whose entries are restricted  by equations (\ref{d}). Thus, equations (\ref{d}) can be interpreted as the defining equations of a codimension two submanifold $X$ in the Grassmannian $ {\bf Gr}(d, V)$. Solutions to system (\ref{d}) correspond to $d$-dimensional submanifolds of 
$V$   whose tangent spaces (translated to the origin) are contained in $X$. Equations of type (\ref{d}) arise in numerous applications in the theory of dispersionless integrable systems, general relativity and  differential geometry. For $d=3$ their integrability aspects, as well as the geometry of the associated fourfolds $X\subset {\bf Gr}(3, 5)$, were 
thoroughly investigated in \cite{DFKN}.

In this paper we consider an important subclass of multi-dimensional ($d\geq 3$)  equations (\ref{d})  known as systems of  Monge-Amp\`ere type,
 \begin{equation}
\begin{array}{c}
a^{ij}(u_iv_j-u_jv_i)+b^iu_i+c^iv_i+m=0, \\
\alpha^{ij}(u_iv_j-u_jv_i)+\beta^iu_i+\gamma^iv_i+\mu=0,
\end{array}
\label{Monge}
\end{equation}
where  each equation corresponds to a constant-coefficient linear combination of the  minors of  $U$. 
Systems of  type (\ref{Monge}) were discussed previously in \cite{Boillat1} from the point of view of `complete exceptionality' of the  Cauchy problem. Geometrically, submanifolds $X$ associated with such systems  are linear sections of the Pl\"ucker embedding of  $ {\bf Gr}(d, V)$ into $\mathbb{P}\Lambda^2(V)$. Note that the class of Monge-Amp\`ere systems is invariant under the natural action of the equivalence group ${\bf SL}(V)$. 
In what follows we assume  systems (\ref{d}), (\ref{Monge})  to be  {\it non-degenerate} in the sense that the corresponding characteristic variety, 
$$
\det \left[\sum_{i=1}^d
p_i \left(\begin{array}{cc}
F_{u_i} & F_{v_i}\\
G_{u_i} & G_{v_i}
\end{array}\right)
\right]=0,
$$
defines an irreducible quadric of rank d for $d\leq 4$, and rank 4 for $d> 4$ (note that 4 is the maximal possible value for the rank of a quadratic form representable as the determinant of a $2\times 2$ matrix with entries linear in $p_i$). This non-degeneracy property holds for all examples of physical/geometric relevance. 

\medskip

Our main results can be summarised as follows:

\begin{itemize}

\item All Monge-Amp\`ere systems (\ref{Monge}) are integrable, with  Lax representations in parameter-dependent commuting vector fields. This result was, in a sense, unexpected: indeed, it was demonstrated in \cite{DF} that  second-order analogues of systems (\ref{d}), known as symplectic Monge-Amp\`ere equations, are not integrable in general for $d\geq 3$.  Our approach is based on the observation that every Monge-Amp\`ere system (\ref{Monge}) can be defined by a pair of differential $d$-forms in $V$, that is, by two elements of $\Lambda^d(V^*)$. Utilising the ${\bf SL}(V)$-equivariant duality between $\Lambda^d(V^*)$ and $\Lambda^2(V)$ we can reduce  the theory of normal forms of Monge-Amp\`ere systems to the  classification of pencils of skew-symmetric two-forms. This, however, is the classical territory (in Sect. \ref{sec:normal} we recall the main ingredients of the theory of Jordan-Kronecker normal forms of skew-symmetric matrix pencils). Thus we obtain normal forms of Monge-Amp\`ere systems in all dimensions $d$ (see below), for which the integrability can be established directly. 

\item For $d=2, 3$ any non-degenerate system  of Monge-Amp\`ere type is linearisable 
(Theorem \ref{MA23} of Sect. \ref{sec:MA23}).

\item For $d=4$ any non-degenerate  system  of Monge-Amp\`ere type is  ${\bf SL}(6)$-equivalent to one of the following  normal forms (Theorem \ref{MA4} of Sect. \ref{sec:MA4}):
\begin{enumerate}
\item $ u_2-v_1=0, ~~~ u_3+v_4=0,$
\item $ u_2-v_1=0, ~~~ u_3+v_4+u_1v_2-u_2v_1=0,$
\item $ u_2-v_1=0, ~~~ u_3v_4-u_4v_3-1=0,$
\item $ u_2-v_1=0, ~~~ u_1+v_2+u_3v_4-u_4v_3=0,$
\end{enumerate}
see Sect. \ref{sec:MA4} for the associated Lax representations. 
Introducing a potential $w$ such that $w_1=u, \ w_2=v,$ one obtains  well-known integrable second-order PDEs: $w_{13}+w_{24}=0$ (linear equation),  
$w_{13}+w_{24}+w_{11}w_{22}-w_{12}^2=0$ (second heavenly equation \cite{Plebanski}), $w_{13}w_{24}-w_{14}w_{23}-1=0$ (first 
heavenly equation \cite{Plebanski}), and $w_{11}+w_{22}+w_{13}w_{24}-w_{14}w_{23}=0$ (Husain equation \cite{Husain}), respectively. All of them originate from self-dual Ricci-flat geometry. 

\item For $d=5$ any non-degenerate system  of Monge-Amp\`ere  type is  ${\bf SL}(7)$-equivalent to one of the following  normal forms (Theorem \ref{MA5} of Sect. \ref{sec:MA5}):
\begin{enumerate}
\item $ u_1+v_2+u_3v_4-u_4v_3 = 0, ~~~ u_2+v_3+u_4v_5-u_5v_4=0,$
\item $ u_2-v_1=0, ~~~ u_1+v_5+u_3v_4-u_4v_3=0,$
\item $ u_2-v_1=0, ~~~ u_4+v_5+u_1v_3-u_3v_1=0,$
\item $ u_2-v_1=0, ~~~ u_5+u_3v_4-u_4v_3=0,$
\end{enumerate}
see Sect. \ref{sec:MA5} for the associated Lax representations. Note that most of the above normal forms  (apart from case 1, $d=5$) can be obtained as travelling wave reductions of the 6-dimensional integrable Monge-Amp\`ere system
\begin{equation}
u_2-v_1=0, ~~~ u_5+v_6+u_3v_4-u_4v_3=0, 
\label{6}
\end{equation}
which reduces to the second-order equation
$w_{15}+w_{26}+w_{13}w_{24}-w_{14}w_{23}=0$ for a potential $w$ defined as $w_1=u, \ w_2=v$. This equation appeared in  hyper-K\"ahler geometry \cite{Takasaki1}  and can be obtained as a reduction of
 $sdiff(\Sigma^2)$  self-dual Yang-Mills equations  \cite{Przanovski}.  

\item For arbitrary  $d$ {\it generic} normal forms are discussed in Sect. \ref{sec:MAd}. 
Note that the cases of even/odd  dimensions   lead to essentially different  normal forms. Thus, for  even  $d=2k+2$ (Jordan case) a  generic Monge-Amp\`ere system can be reduced to the form
$$
\begin{array}{c}
u_{2k+1}=(u_1v_2-u_2v_1) +(u_3v_4-u_4v_3) + ... +(u_{2k-1}v_{2k} - u_{2k}v_{2k-1}),\\
v_{2k+2}= a_1 (u_1v_2-u_2v_1) + a_2(u_3v_4-u_4v_3) + ... + a_k  (u_{2k-1}v_{2k} - u_{2k}v_{2k-1}),
\end{array}
$$
here $a_1, \dots, a_k$ are arbitrary constants. For  odd  $d=2k+1$ (Kronecker case) a  generic Monge-Amp\`ere system can be reduced to the form
$$
\begin{array}{c}
u_1 + v_2 =(u_3v_4 - u_4v_3) + (u_5v_6 - u_6v_5) + ...+ (u_{2k-1}v_{2k} - u_{2k}v_{2k-1}), \\
u_2 + v_3 = (u_4v_5 - u_5v_4)+ (u_6v_7 - u_7v_6) + ... + (u_{2k}v_{2k+1} - u_{2k+1}v_{2k}),
\end{array}
$$
see Sect. \ref{sec:MAd} for the associated Lax representations.


\item One can show that all Monge-Amp\`ere systems of type (\ref{Monge}) possess infinitely many hydrodynamic reductions, see \cite{Fer22, Fer3} for further details. 

\item In Theorem \ref{MAchar} of Sect. \ref{sec:charMA}  we demonstrate that the necessary and sufficient conditions for a codimension two submanifold $X\subset {\bf Gr}(d, V)$ to be a linear section is that the only `essential' second fundamental forms of $X$ are the ones coming from $ {\bf Gr}(d, V)$ itself. This property can be reformulated as a system of second-order differential constraints for the functions $F, G$ defining system (\ref{d}) thus providing an invariant differential-geometric characterisation of  Monge-Amp\`ere systems.

\end{itemize}

\section{Classification of Monge-Amp\`ere systems}
\label{sec:MA}


\subsection{Jordan-Kronecker normal forms of skew-symmetric pencils}
\label{sec:normal}

Here we follow \cite{Gauger} to review Jordan-Kronecker normal forms  of skew-symmetric pencils on a vector space $V$ of dimension $d+2$. Any such pencil gives rise to two elements in $\Lambda^2(V)$. Taking the dual elements in $\Lambda^d(V^*)$ and equating them to zero we obtain normal forms of  Monge-Amp\`ere systems. 

A skew-symmetric pencil can be  written in the form $\mu A+
\lambda B$ where $A$ and $B$ are skew-symmetric matrices considered modulo simultaneous transformations $A\to XAX^t, \ B\to XBX^t$,  while $[\lambda : \mu]\in \mathbb{P}^1$ is defined modulo automorphisms of $\mathbb{P}^1$. Normal forms of such
pencils are classified by the following data:
\begin{itemize}
\item minimal indices $0\le m_1\le m_2\le \dots \le m_p$, \ $p\ge0$ (in particular, the set of minimal indices can be empty). Each minimal index $m_i$ corresponds to a Kronecker block $\cM_{m_i}$ of the odd size $(2m_i+1)\times(2m_i+1)$.
\item elementary divisors $(a_1\mu+b_1\lambda)^{e_1}$, \dots,
$(a_q\mu+b_q\lambda)^{e_q}$ 
where $[a_i:b_i]$ are considered as points in $\mathbb{P}^1$. Each elementary divisor 
$(a_i\mu+b_i\lambda)^{e_i}$ corresponds to a Jordan block $\cE_{e_i}[a_i:b_i]$ of the even size $2e_i\times 2e_i$.
\end{itemize}
Explicitly, the canonical form of the pencil specified by  these data is
\begin{equation}\label{eq1}
P = \begin{pmatrix} \cM_{m_1} & & & & & \\
 & \ddots & & & & \\
 & & \cM_{m_p} & & & \\
 & & & \cE_{e_1}[a_1:b_1] & & \\
 & & & & \ddots & \\
 & & & & & \cE_{e_q}[a_q:b_q] \\
\end{pmatrix}
\end{equation}
where the Kronecker blocks $\cM_m$ and the Jordan blocks $\cE_{n}[a:b]$ are defined as follows:
\begin{align*}
\cM_m & = \begin{pmatrix} 0 & M_m \\
-M_m^t & 0
\end{pmatrix},
\quad {\rm size} \ (2m+1)\times(2m+1), \  \cM_0 = (0),\\
\ \\
\cE_n([1:b]) &= \begin{pmatrix} 0 & E_n(b) \\
-E_n(b)^t & 0
\end{pmatrix}, 
\quad {\rm size} \ (2n)\times(2n),\\
\ \\
\cE_n([0:1]) &= \begin{pmatrix} 0 & F_n \\
-F_n^t & 0
\end{pmatrix},
\quad {\rm size} \ (2n)\times(2n).\\
\end{align*}
Here we use the notation
\begin{align*}
M_m &= \begin{pmatrix}
&&&&&& \lambda \\
&&&&& \lambda & \mu \\
&&&&\cdot &\mu & \\
&&& \cdot &\cdot && \\
&& \cdot &\cdot &&& \\
& \lambda & \cdot &&&& \\
\lambda & \mu &&&&& \\
\mu &&&&&&
\end{pmatrix},
\quad {\rm size} \ (m+1)\times m,\\
\ \\
E_n(b) &=
\begin{pmatrix}
&&&&& \mu + b\lambda \\
&&&& \cdot & \lambda \\
&&& \cdot && \\
&& \cdot && \cdot & \\
&&& \cdot && \\
& \mu+b\lambda & \cdot &&& \\
\mu+b\lambda & \lambda  &&&&
\end{pmatrix},
\quad {\rm size} \ n \times n,\\
\ \\
F_n &=
\begin{pmatrix}
&&&&& \lambda \\
&&&& \cdot & \mu \\
&&& \cdot && \\
&& \cdot && \cdot & \\
&&& \cdot && \\
& \lambda & \cdot &&& \\
\lambda & \mu  &&&&
\end{pmatrix},
\quad {\rm size} \ n \times n.
\end{align*}
In addition,  elementary divisors are considered up to non-degenerate linear
transformations of $\lambda$ and $ \mu$,  in other words, parameters $[a_i:b_i]$ are considered modulo projective transformations.
We also impose the following non-degeneracy conditions:
\begin{itemize}
\item The pencil does not have zero minimal indices (that is, no $1\times 1$ zero Kronecker blocks $\cM_0$). Otherwise, the corresponding Monge-Amp\`ere system  reduces to a system of lower dimension.
\item For $d\geq 3$, the pencil does not contain elements of rank two. These elements correspond to equations of the type $u_i=0$ and result in degenerate systems with characteristic varieties of rank 2.
\end{itemize}

Any  element of rank four in the pencil gives rise to the equation of the type $u_2-v_1=0$. Introducing the potential $w$ such that $w_1=u$, $w_2=v$, we can reduce the corresponding system to a single second-order Monge-Amp\`ere equation for $w$. Note  that a pencil may  contain several  elements of rank four that might lead to non-equivalent second-order Monge-Amp\`ere equations (see  Remark 1 in Sect. \ref{sec:MA4}).

\subsection{Linearisability of Monge-Amp\`ere systems for $d=2, 3$}
\label{sec:MA23}

The classification of $4\times 4$ and $5\times 5$ skew-symmetric pencils leads to the following result:

\begin{theorem}  \label{MA23}
For $d=2, 3$, any non-degenerate  system  of Monge-Amp\`ere  type is  linearisable.
\end{theorem}

\medskip

\centerline{\bf Proof:}

For $d=2$ one needs to classify non-degenerate $4\times 4$ pencils. Note that in this case we allow elements of rank two in the pencil. There are only two non-equivalent normal forms without zero minimal indices, namely
\[
\left(\begin{smallmatrix}
. & \mu & . & .  \\
-\mu & . & . &  .  \\
. & . & . & \lambda  \\
. & . & -\lambda & . 
\end{smallmatrix}\right),\quad
\left(\begin{smallmatrix}
. & . & . & \lambda &  \\
. & . & \lambda &  \mu &  \\
. & -\lambda & . & .  \\
-\lambda & -\mu & . & . 
\end{smallmatrix}\right).
\]
Both pencils give rise to linear systems. Indeed, the first pencil corresponds to  2-forms
$$
dz^1\wedge dz^2 ~~~ {\rm and} ~~~ dz^3\wedge dz^4.
$$
($z^1, \dots, z^4$ denote coordinates in 4-dimensional space $V$). Setting $u=z^4, \ v=z^2,  \ x^1=z^1,\ x^2=z^3$ and equating these 2-forms to zero we obtain the linear hyperbolic system $u_1=0, \ v_2=0$ (note that we do not need to use the duality transformation  for $d=2$). Similarly, the second pencil corresponds to 2-forms
$$
dz^1\wedge dz^4+dz^2\wedge dz^3 ~~~ {\rm and} ~~~ dz^2\wedge dz^4.
$$
Setting $u=z^4, \ v=z^3, \ x^1=z^1,\ x^2=z^2$ and equating these 2-forms to zero we obtain the linear parabolic system $u_1=0, \ v_1-u_2=0$.

For $d=3$ one needs to classify non-degenerate $5\times 5$ pencils. The non-degeneracy constraints imply that the only possibility is a single $5\times 5$ Kronecker block,
\[
\begin{pmatrix}
. & . & . & . & \lambda  \\
. & . & . & \lambda & \mu  \\
. & . & . & \mu & . \\
. & -\lambda & -\mu & . & .  \\
-\lambda & -\mu & . & . & . \\
\end{pmatrix}.
\]
It is generated by  the bi-vectors
$$
\partial_{z^1}\wedge \partial_{z^5}+\partial_{z^2}\wedge \partial_{z^4} ~~~ {\rm and} ~~~ \partial_{z^2}\wedge \partial_{z^5}+\partial_{z^3}\wedge \partial_{z^4},
$$
the corresponding dual 3-forms are 
$$
dz^2\wedge dz^3\wedge dz^4+dz^1\wedge dz^3\wedge dz^5, ~~~ 
dz^1\wedge dz^3\wedge dz^4+dz^1\wedge dz^2\wedge dz^5.
$$
Setting $u=z^5, \ v=z^4, \ x^1=z^1,\ x^2=z^2,\ x^3=z^3$ and equating these 3-forms to zero we obtain the linear hyperbolic system $v_1-u_2=0, \ v_2-u_3=0$. This finishes the proof of Theorem \ref{MA23}.

\medskip

We  emphasize that the linearisability of Monge-Amp\`ere systems for $d=2, 3$ does not generalise to higher dimensions $d\geq 4$, see the classification results below.

\subsection{Classification of Monge-Amp\`ere systems for $d=4$}
\label{sec:MA4}

The classification of $6\times 6$ skew-symmetric pencils leads to the following result:

\begin{theorem}  \label{MA4}
 In four dimensions, any non-degenerate  system  of Monge-Amp\`ere  type is  ${\bf SL}(6)$-equivalent to one of the following  normal
forms:
\begin{enumerate}
\item $ u_2-v_1=0, ~~~ u_3+v_4=0,$
\item $ u_2-v_1=0, ~~~ u_3+v_4+u_1v_2-u_2v_1=0,$
\item $ u_2-v_1=0, ~~~ u_3v_4-u_4v_3-1=0,$
\item $ u_2-v_1=0, ~~~ u_1+v_2+u_3v_4-u_4v_3=0.$
\end{enumerate}

\end{theorem}

\medskip

\centerline{\bf Proof:}

 One needs to classify non-degenerate $6\times 6$  skew-symmetric pencils. First assume that there is a non-empty set of minimal indices. As any minimal index of the pencil corresponds to a Kronecker block of odd size, there should be two of them, both equal to 1. This leads to the normal form consisting of two $3\times 3$ Kronecker blocks,
\[
\begin{pmatrix}
. & . & \lambda & . & . & . \\
. & . & \mu &  . & . & . \\
-\lambda & -\mu & . & . & . & . \\
. & . & . & . & . & \lambda\\
. & . & . & . & . & \mu \\
. & . & . & -\lambda & -\mu &  . 
\end{pmatrix},
\]
which corresponds to  linear system 1. Indeed, the above pencil is generated by  the bi-vectors
$$
\partial_{z^1}\wedge \partial_{z^3}+\partial_{z^4}\wedge \partial_{z^6} ~~~ {\rm and} ~~~ \partial_{z^2}\wedge \partial_{z^3}+\partial_{z^5}\wedge \partial_{z^6}.
$$
The corresponding dual 4-forms are
$$
dz^2\wedge dz^4\wedge dz^5\wedge dz^6+dz^1\wedge dz^2\wedge dz^3\wedge dz^5, ~~~ 
dz^1\wedge dz^4\wedge dz^5\wedge dz^6+dz^1\wedge dz^2\wedge dz^3\wedge dz^4.
$$
Setting $u=z^6, \ v=z^3, \ x^1=z^4,\ x^2=z^1,\ x^3=-z^2,\  x^4=z^5$ and equating these 4-forms to zero we obtain  linear system 1. 

Now assume that the set of minimal indices is empty. The non-degeneracy assumption implies that for any  $[a_i:b_i]$, there can be only one elementary divisor $(a_i\mu+b_i\lambda)^{e_i}$. So, up to projective transformations the only possible lists of elementary divisors are:
\begin{itemize}
\item $\{\lambda^3\}$,
\item $\{\lambda^2, \ \mu\}$,
\item $\{\lambda, \ \mu, \ \lambda+\mu\}$.
\end{itemize}
Explicitly, the associated pencils have the form
\[
\left(\begin{smallmatrix}
. & . & . & . & . & \lambda \\
. & . & . &  . & \lambda & \mu \\
. & . & . & \lambda & \mu & . \\
. & . & -\lambda & . & . & . \\
. & -\lambda & -\mu & . & . & . \\
-\lambda & -\mu & . & . &  . &  . 
\end{smallmatrix}\right),\quad
\left(\begin{smallmatrix}
. & . & . & \lambda & . & . \\
. & . & \lambda &  \mu & . & . \\
. & -\lambda & . & . & . & . \\
-\lambda & -\mu & . & . & . & . \\
. & . & . & . & . & \mu \\
. & . & . & . & -\mu &  . 
\end{smallmatrix}\right),\quad
\left(\begin{smallmatrix}
. & \lambda & . & . & . & . \\
-\lambda & . & . &  . & . & . \\
. & . & . & \mu & . & . \\
. & . & -\mu & . & . & . \\
. & . & . & . & . & \lambda+\mu \\
. & . & . & . & -\lambda-\mu &  . 
\end{smallmatrix}\right),
\]
which correspond to  systems 2-4, respectively.  This finishes the proof of Theorem \ref{MA4}. 

\bigskip

\noindent {\bf Remark 1.} Let us consider  system 3,
$$ 
u_2-v_1=0, ~~~ u_3v_4-u_4v_3-1=0,
$$
which is related to the first heavenly equation. Interchanging the roles of $u$ and $x^3$ we obtain the equivalent system,
$$
 u_3-v_4=0, ~~~ u_2+v_1u_3-v_3u_1=0,
$$
which leads to the modified heavenly equation, $w_{24}+w_{13}w_{34}-w_{33}w_{14}=0$, for the potential $w$ defined by the relations
  $w_4=u, \ w_3=v$. The modified heavenly equation appeared recently in the classification of  integrable  symplectic Monge-Amp\`ere equations  \cite{DF}. Thus, system 3 provides a B\"acklund transformation connecting the first heavenly and the modified heavenly equations. We point out that these second-order equations are not equivalent under the natural equivalence group ${\bf Sp}(8)$ acting on symplectic Monge-Amp\`ere equations in 4D. 

\medskip

\noindent {\bf Remark 2.} All  nonlinear systems from Theorem \ref{MA4} possess Lax pairs of the form $[X, Y]=0$ where $X$ and $Y$ are parameter-dependent vector fields. 

\noindent {\it System 2:} $ u_2-v_1=0, ~~ u_3+v_4+u_1v_2-u_2v_1=0,$
$$
X=\partial_4+u_{1}\partial_2-u_{2}\partial_1+\lambda \partial_1, ~~~
Y=\partial_3-v_{1}\partial_2+v_{2}\partial_1-\lambda \partial_2.
$$
\noindent {\it System 3:} $ u_2-v_1=0, ~~ u_{3}v_{4}-u_{4}v_{3}-1=0$,
$$
X=u_{3}\partial_4-u_{4}\partial_3+\lambda \partial_1, ~~~
Y=-v_{3}\partial_4+v_{4}\partial_3-\lambda \partial_2.
$$
\noindent {\it System 4:} $ u_2-v_1=0, ~~ u_{1}+v_{2}+u_{3}v_{4}-u_{4}v_{3}=0$, 
$$
X=\partial_2+u_{3}\partial_4-u_{4}\partial_3+\lambda \partial_1, ~~~
Y=\partial_1-v_{3}\partial_4+v_{4}\partial_3-\lambda \partial_2.
$$
 Modifications
of the inverse scattering transform and  the  $\overline\partial$-dressing method
for Lax pairs of this type were developed in \cite{Manakov, Manakov3,
Bogdanov2}.

\subsection{Classification of Monge-Amp\`ere  systems for $d=5$}
\label{sec:MA5}

The classification of $7\times 7$ skew-symmetric pencils leads to the following result:

\begin{theorem}\label{MA5}
In five dimensions, any non-degenerate system  of Monge-Amp\`ere  type is  ${\bf SL}(7)$-equivalent to one of the following  normal
forms:
\begin{enumerate}
\item $ u_1+v_2+u_3v_4-u_4v_3 = 0, ~~~ u_2+v_3+u_4v_5-u_5v_4=0,$
\item $ u_2-v_1=0, ~~~ u_1+v_5+u_3v_4-u_4v_3=0,$
\item $ u_2-v_1=0, ~~~ u_4+v_5+u_1v_3-u_3v_1=0,$
\item $ u_2-v_1=0, ~~~ u_5+u_3v_4-u_4v_3=0.$
\end{enumerate}
\end{theorem}

\medskip

\centerline{\bf Proof:}

\medskip
One  needs to classify  non-degenerate  $7 \times 7$ skew-symmetric pencils. As the size of matrices is odd, the set of minimal indices cannot be empty. Simple analysis shows that there can be at most one minimal index equal to 1, 2 or 3. The latter case is generic and corresponds to the single Kronecker block
\[
\left(\begin{smallmatrix}
. & . & . & . & . & . & \lambda \\
. & . & . & . & . & \lambda & \mu \\
. & . & . & . & \lambda & \mu & . \\
. & . & . & . & \mu & . & . \\
. & . & -\lambda & -\mu & . & . & . \\
. & -\lambda & -\mu & . & . & . & . \\
-\lambda & -\mu & . & . & . & . & . 
\end{smallmatrix}\right).
\]
It leads to system 1. If the minimal index is 2, then we can assume that the remaining elementary divisor is $\lambda$. If the minimal index is 1, then the possible lists of minimal divisors are equivalent to $\{ \lambda^2 \}$ or $\{\lambda, \mu \}$. Explicitly, these three pencils are:
\[
\left(\begin{smallmatrix}
 . & . & . & . & \lambda & . & . \\
 . & . & . & \lambda & \mu & . & . \\
. & . & . & \mu & . & . & . \\
. & -\lambda & -\mu & . & . & . & . \\
-\lambda & -\mu & . & . & . & . & . \\
. & . & . & . & . & . & \lambda \\
. & . & . & . & . & -\lambda & . \\
\end{smallmatrix}\right),\quad
\left(\begin{smallmatrix}
. & . & \lambda & . & . & . & . \\
. & . & \mu & . & . & . & . \\
-\lambda & -\mu & . & . & . & . & . \\
. & . & . & . & . & . & \lambda \\
. & . & . & . & . & \lambda & \mu \\
. & . & . & . & -\lambda & . & . \\
. & . & . & -\lambda & -\mu & . & . 
\end{smallmatrix}\right),\quad 
\left(\begin{smallmatrix}
. & . & \lambda & . & . & . & . \\
. & . & \mu & . & . & . & . \\
-\lambda & -\mu & . & . & . & . & . \\
. & . & . & . & \lambda & . & . \\
. & . & . & -\lambda & . & . & . \\
. & . & . & . & . & . & \mu \\
. & . & . & . & . & -\mu & . 
\end{smallmatrix}\right).
\]
The corresponding systems are 2, 3 and 4, respectively. 
This finishes the proof of Theorem \ref{MA5}.

\medskip

\noindent {\bf Remark.} All  systems from Theorem \ref{MA5} possess Lax pairs of the form $[X, Y]=0$ where $X$ and $Y$ are parameter-dependent vector fields. 

\noindent {\it System 1:}  $ u_1+v_2+u_3v_4-u_4v_3 = 0, ~~~ u_2+v_3+u_4v_5-u_5v_4=0,$
$$
\begin{array}{c}
X=\partial_2+\lambda \partial_3+u_{3}\partial_4-u_{4}\partial_3+\lambda (u_4\partial_5-u_5\partial_4), \\
Y=\partial_1+\lambda \partial_2-v_{3}\partial_4+v_{4}\partial_3-\lambda (v_4\partial_5-v_5\partial_4).
\end{array}
$$
\noindent {\it System 2:} $ u_2-v_1=0, ~~~ u_1+v_5+u_3v_4-u_4v_3=0,$
$$
X=\partial_5+u_{3}\partial_4-u_{4}\partial_3+\lambda \partial_1, ~~~
Y=\partial_1-v_{3}\partial_4+v_{4}\partial_3-\lambda \partial_2.
$$
\noindent {\it System 3:}  $ u_2-v_1=0, ~~~ u_4+v_5+u_1v_3-u_3v_1=0,$
$$
X=\partial_5+u_{1}\partial_3-u_{3}\partial_1+\lambda \partial_1, ~~~
Y=\partial_4-v_{1}\partial_3+v_{3}\partial_1-\lambda \partial_2.
$$
\noindent {\it System 4:} $ u_2-v_1=0, ~~~ u_5+u_3v_4-u_4v_3=0,$
$$
X=u_{3}\partial_4-u_{4}\partial_3+\lambda \partial_1, ~~~
Y=\partial_5-v_{3}\partial_4+v_{4}\partial_3-\lambda \partial_2.
$$

\subsection{Monge-Amp\`ere systems for arbitrary $d$ }
\label{sec:MAd}

Since  normal forms of  skew-symmetric pencils in  even/odd dimensions are essentially different, we will consider these cases separately. Moreover, we will only discuss {\it generic} normal forms.

\medskip

\noindent {\bf Even dimension.} For $d=2k+2$ a  generic skew-symmetric pencil can be brought to the Jordan normal form with $2\times 2$ blocks along the diagonal. The corresponding  system is
$$
\begin{array}{c}
u_{2k+1}=(u_1v_2-u_2v_1) +(u_3v_4-u_4v_3) + ... +(u_{2k-1}v_{2k} - u_{2k}v_{2k-1}),\\
v_{2k+2}= a_1 (u_1v_2-u_2v_1) + a_2(u_3v_4-u_4v_3) + ... + a_k  (u_{2k-1}v_{2k} - u_{2k}v_{2k-1}),
\end{array}
$$
here $a_1, \dots, a_k$ are arbitrary constants. Relabelling coordinates we can rewrite these equations in the form
$$
u_{t}=\sum_{i=1}^k(u_{x^i}v_{y^i}-u_{y^i}v_{x^i}) , ~~~
v_{\tau}=\sum_{i=1}^ka_i(u_{x^i}v_{y^i}-u_{y^i}v_{x^i}). 
$$
The corresponding Lax pair is given by
$$
X=\partial_{\tau}+\sum_{i=1}^k\alpha_i(u_{x^i}\partial_{y^i}-u_{y^i}\partial_{x^i}), ~~~ Y=\partial_{t}+\sum_{i=1}^k\beta_i(v_{x^i}\partial_{y^i}-v_{y^i}\partial_{x^i}),
$$
where $\alpha_i=\frac{1}{\lambda}-a_i, \ \beta_i=1-\lambda a_i$.

\medskip

\noindent {\bf Odd dimension.} For  $d=2k+1$ a  generic skew-symmetric pencil can be brought to the Kronecker normal form. The corresponding system is
$$
\begin{array}{c}
u_1 + v_2 =(u_3v_4 - u_4v_3) + (u_5v_6 - u_6v_5) + ...+ (u_{2k-1}v_{2k} - u_{2k}v_{2k-1}), \\
u_2 + v_3 = (u_4v_5 - u_5v_4)+ (u_6v_7 - u_7v_6) + ... + (u_{2k}v_{2k+1} - u_{2k+1}v_{2k}).
\end{array}
$$
Its Lax pair is given by
$$
X=\partial_2+\lambda \partial_3-\sum_{i=3}^{2k}\alpha_i (u_{i}\partial_{i+1}-u_{i+1}\partial_i), ~~~
Y=\partial_1+\lambda \partial_2+\sum_{i=3}^{2k}\alpha_i (v_{i}\partial_{i+1}-v_{i+1}\partial_i),
$$
where $\alpha_{2s-1}=1, \ \alpha_{2s}=\lambda$.


\medskip

Since generic normal forms are integrable in any dimension, and integrability is preserved in the limit, we conclude that all  systems of Monge-Amp\`ere type must be integrable.

\section{Differential geometry of  Monge-Amp\`ere systems}
\label{sec:charMA}

Consider   system  (\ref{d}) of dimension $d=m+1$. Representing it in  evolutionary form, 
\begin{equation}
u_{m+1}=f(u_1, \dots, u_m, v_1, \dots, v_m), ~~~ v_{m+1}=g(u_1, \dots, u_m, v_1, \dots, v_m), 
\label{m00}
\end{equation}
we will derive differential constraints for the functions $f$ and $ g$ that characterise systems (\ref{Monge}) of Monge-Amp\`ere type. 
Let us begin with the simplest case $d=2$,
\begin{equation}
u_{2}=f(u_1, v_1), ~~~ v_{2}=g(u_1, v_1), 
\label{m1}
\end{equation}
which however contains all essential ingredients of the general case.

\medskip

\noindent {\bf Proposition 1.} {\it System (\ref{m1}) is of Monge-Amp\`ere type if and only if the (symmetric) differentials $d^2f$ and $d^2g$ are proportional to the quadratic form $dfdv_1-dgdu_1$:
\begin{equation}
d^2f, \ d^2g \  \in \ {\rm span}  \{dfdv_1-dgdu_1\}.
\label{m0}
\end{equation}
}

\centerline {\bf Proof:}

\medskip

\noindent Equations (\ref{m1}) specify a surface $X$ in the Grassmannian ${\bf Gr}(2, 4)$. The Pl\"ucker embedding of ${\bf Gr}(2, 4)$ into $\mathbb{P}^5$  is a quadric with  position vector
$(u_1, \ v_1, \ u_2, \ v_2, \ u_2v_1-u_1v_2).$ 
The induced embedding of $X$ has position vector
$$
R=(u_1, \ v_1, \ f, \ g, \ v_1f-u_1g).
$$ 
To prove that system (\ref{m1}) is of Monge-Amp\`ere type we need to show that  components of $R$ satisfy 2 linear relations with constant coefficients or, equivalently, that the Pl\"ucker image of $X$ lies in a 3-dimensional linear subspace of $\mathbb{P}^5$. This means that the 
union of all osculating spaces of $X$ must be 3-dimensional. Since the tangent space of $X$,  spanned by the vectors
$$
R_{u_1}=(1, \ 0, \ f_{u_1}, \ g_{u_1}, \ v_1f_{u_1}-u_1g_{u_1}-g),
$$
$$
R_{v_1}=(0, \ 1, \ f_{v_1}, \ g_{v_1}, \ v_1f_{v_1}-u_1g_{v_1}+f),
$$
is already 2-dimensional, we have to show that the union of the second- and third-order osculating spaces (spanned by the second- and third-order partial derivatives of the position vector $R$ with respect to $u_1$ and $v_1$) has dimension 1. As higher-order derivatives of $R$ have zeros in the first two positions, the rank of the following matrix must  equal 1:
$$
rk \left(\begin{array}{ccc}
f_{u_1u_1} &  g_{u_1u_1} & v_1f_{u_1u_1}-u_1g_{u_1u_1}-2g_{u_1}\\
f_{u_1v_1} &  g_{u_1v_1} & v_1f_{u_1v_1}-u_1g_{u_1v_1}+f_{u_1}-g_{v_1}\\
f_{v_1v_1} &  g_{v_1v_1} & v_1f_{v_1v_1}-u_1g_{v_1v_1}+2f_{v_1}\\
f_{u_1u_1u_1} &  g_{u_1u_1u_1} & v_1f_{u_1u_1u_1}-u_1g_{u_1u_1u_1}-3g_{u_1}\\
f_{u_1u_1v_1} &  g_{u_1u_1v_1} & v_1f_{u_1u_1v_1}-u_1g_{u_1u_1v_1}+f_{u_1u_1}-2g_{u_1v_1}\\
f_{u_1v_1v_1} &  g_{u_1v_1v_1} & v_1f_{u_1v_1v_1}-u_1g_{u_1v_1v_1}+2f_{u_1v_1}-g_{v_1v_1}\\
f_{v_1v_1v_1} &  g_{v_1v_1v_1} & v_1f_{v_1v_1v_1}-u_1g_{v_1v_1v_1}+3f_{v_1v_1}\\
\end{array}\right)=1.
$$
Since the terms of the third column  containing multiples of $v_1$ or $u_1$ are proportional to the first and second columns, respectively, and can therefore be eliminated without changing the rank, we obtain a simpler condition,
$$
rk \left(\begin{array}{ccc}
f_{u_1u_1} &  g_{u_1u_1} & -2g_{u_1}\\
f_{u_1v_1} &  g_{u_1v_1} & f_{u_1}-g_{v_1}\\
f_{v_1v_1} &  g_{v_1v_1} & 2f_{v_1}\\
f_{u_1u_1u_1} &  g_{u_1u_1u_1} & -3g_{u_1}\\
f_{u_1u_1v_1} &  g_{u_1u_1v_1} & f_{u_1u_1}-2g_{u_1v_1}\\
f_{u_1v_1v_1} &  g_{u_1v_1v_1} & 2f_{u_1v_1}-g_{v_1v_1}\\
f_{v_1v_1v_1} &  g_{v_1v_1v_1} & 3f_{v_1v_1}\\
\end{array}\right)=1.
$$
This condition  is equivalent to the requirement that the first and second columns are proportional to the third column. Let $p$ and $q$ be the corresponding  coefficients of proportionality.  In compact form, this can be represented as
\begin{equation}
d^2f=2p(dfdv_1-dgdu_1), ~~~ d^2g=2q(dfdv_1-dgdu_1),
\label{m2}
\end{equation}
and
\begin{equation}
d^3f=3p(d^2fdv_1-d^2gdu_1), ~~~ d^3g=3q(d^2fdv_1-d^2gdu_1),
\label{m3}
\end{equation}
respectively. Calculating (symmetric) differentials of (\ref{m2}) and comparing the result with  (\ref{m3}) we obtain the equations for $p$ and $q$, 
\begin{equation}
dp=p(pdv_1-qdu_1), ~~~ dq=q(pdv_1-qdu_1).
\label{m4}
\end{equation}
Equations (\ref{m2}) and (\ref{m4}) constitute a closed involutive differential system for $f$ and $g$ which characterises Monge-Amp\`ere systems. It remains to note that conditions (\ref{m4}) can be obtained as the consistency conditions of  equations (\ref{m2}) alone, without using (\ref{m3}). In other words, equations (\ref{m2}) imply both (\ref{m3}) and (\ref{m4}). This finishes the proof of Proposition 1. 

\medskip 

\noindent {\bf Remark 1.} Condition (\ref{m0}) has a clear projective-geometric interpretation. Recall that  the second fundamental forms of $X\subset \mathbb{P}^5$ are spanned by $d^2f, \ d^2g$ and $dfdv_1-dgdu_1$. Here the last form is the restriction to $X$ of the second fundamental form of the Grassmannian ${\bf Gr}(2, 4)$, namely, $du_2dv_1-dv_2du_1$. Thus, (\ref{m0}) says that the only `essential' second fundamental form of $X\subset  {\bf Gr}(2, 4)$ is the one coming from the second fundamental form of  ${\bf Gr}(2, 4)\subset \mathbb{P}^5$. This property is clearly necessary   for $X$ to be a linear section. The above result shows that in this particular case it is also sufficient.

\medskip 

\noindent {\bf Remark 2.} Condition (\ref{m0}) can be written as a system of PDEs for $f$ and $g$, indeed, the elimination of $p$ and $q$ from (\ref{m2})  implies the second-order relations
\begin{equation}
\begin{array}{c}
f_{u_1u_1}=\frac{2g_{u_1}}{g_{v_1}-f_{u_1}}f_{u_1v_1}, ~~~ f_{v_1v_1}=\frac{2f_{v_1}}{f_{u_1}-g_{v_1}}f_{u_1v_1}, \\
\ \\
g_{u_1u_1}=\frac{2g_{u_1}}{g_{v_1}-f_{u_1}}g_{u_1v_1}, ~~~ g_{v_1v_1}=\frac{2f_{v_1}}{f_{u_1}-g_{v_1}}g_{u_1v_1}. 
\end{array}
\label{fg1}
\end{equation}

\medskip

The case of arbitrary dimension $d=m+1$ can be considered in a similar way.

\medskip

\begin{theorem} \label{MAchar} System (\ref{m00}) is of Monge-Amp\`ere type if and only if 
\begin{equation}
d^2f, \ d^2g \  \in \ {\rm span}  \{du_idv_j-du_jdv_i, \ dfdv_i-dgdu_i\  \vert \ i, j=1,\dots, m\}.
\label{m000}
\end{equation}
\end{theorem}

\centerline {\bf Proof:}

\medskip

\noindent  Equations (\ref{m00}) specify a submanifold $X$ in  ${\bf Gr}(d, V)$ whose Pl\"ucker embedding into $\mathbb{P}\Lambda^2(V)$ has  position vector
$$
(u_i, \ v_i, \ f, \ g,  \ u_iv_j-u_jv_i, \ v_if-u_ig), ~~~ i, j=1,\dots, m.
$$
To prove that system (\ref{m00}) is of Monge-Amp\`ere type we need to show that  $X$ lies in a  linear subspace of codimension two.  Calculation of osculating spaces similar to that from the proof of Proposition 1 implies that this requirement is equivalent to the  conditions
\begin{equation}
\begin{array}{c}
d^2f=2a^{ij}(du_idv_j-du_jdv_i)+2p^i(dfdv_i-dgdu_i), \\ 
d^2g=2b^{ij}(du_idv_j-du_jdv_i)+2q^i(dfdv_i-dgdu_i),
\end{array}
\label{m5}
\end{equation}
as well as
\begin{equation}
d^3f=3p^i(d^2fdv_i-d^2gdu_i), ~~~ d^3g=3q^i(d^2fdv_i-d^2gdu_i),
\label{m6}
\end{equation}
(the standard summation convention is assumed). Moreover, for the first two terms in (\ref{m5}) we assume $i<j$.  Calculating (symmetric) differentials of (\ref{m5}) and comparing the result with  (\ref{m6}) we obtain the equations for the coefficients, 
\begin{equation}
\begin{array}{c}
da^{ij}=a^{ij}\omega_1-b^{ij}\omega_2, ~~~ db^{ij}=a^{ij}\varphi_1-b^{ij}\varphi_2, \\
dp^i=p^i\omega_1-q^i\omega_2, ~~~ dq^i=p^i\varphi_1-q^i\varphi_2,
\end{array}
\label{m7}
\end{equation}
where we adopt the notation
$$
\omega_1=p^idv_i, ~~~ \omega_2=p^idu_i, ~~~ \varphi_1=q^idv_i, ~~~ \varphi_2=q^idu_i.
$$
We point out that, modulo (\ref{m7}), these forms satisfy the structure equations
$$
d\omega_1=\varphi_1\wedge \omega_2, ~~~ d\omega_2=\omega_1\wedge \omega_2+\varphi_2\wedge\omega_2, ~~~
d\varphi_1=\varphi_1\wedge \varphi_2+\varphi_1\wedge\omega_1, ~~~ d\varphi_2=\varphi_1\wedge \omega_2.
$$
Equations (\ref{m5}) and (\ref{m7}) constitute a closed involutive differential system for $f$ and $g$ which characterises Monge-Amp\`ere systems. It remains to point out that conditions (\ref{m7}) can be obtained as the consistency conditions of  equations (\ref{m5}) alone, without using (\ref{m6}). In other words, equations (\ref{m5}) imply both (\ref{m6}) and (\ref{m7}). This finishes the proof of Theorem \ref{MAchar}. 

\medskip

\noindent{\bf Remark 3.} Condition (\ref{m000}) means that the only essential second fundamental forms of the submanifold $X\subset {\bf Gr}(d, V)$ are the ones coming from the Grassmannian itself. These conditions can be written as a system of second-order PDEs for $f$ and $g$,
\begin{equation}
\begin{array}{c}
f_{u_iu_i}=\frac{2g_{u_i}}{g_{v_i}-f_{u_i}}f_{u_iv_i}, ~~~ f_{v_iv_i}=\frac{2f_{v_i}}{f_{u_i}-g_{v_i}}f_{u_iv_i}, \\
\ \\
f_{u_iu_j}=\frac{g_{u_j}}{g_{v_i}-f_{u_i}}f_{u_iv_i}+\frac{g_{u_i}}{g_{v_j}-f_{u_j}}f_{u_jv_j}, ~~~ 
f_{v_iv_j}=\frac{f_{v_j}}{f_{u_i}-g_{v_i}}f_{u_iv_i}+\frac{f_{v_i}}{f_{u_j}-g_{v_j}}f_{u_jv_j}, \\
\ \\
f_{u_iv_j}+f_{u_jv_i}=\frac{f_{u_j}-g_{v_j}}{f_{u_i}-g_{v_i}}f_{u_iv_i}+
\frac{f_{u_i}-g_{v_i}}{f_{u_j}-g_{v_j}}f_{u_jv_j};
\end{array}
\label{fij}
\end{equation}
here $i, j$ take any values from 1 to m; the equations for $g$ can be obtained by the simultaneous substitution $f\leftrightarrow g$ and $u\leftrightarrow v$. For $m=1$ these conditions  reduce to (\ref{fg1}).

\medskip

\noindent {\bf {Remark 4.}} Each equation (\ref{fij}) involves maximum two distinct indices, namely $i$ and $j$. Thus, if all traveling wave reductions of  system (\ref{m00}) to 3D obtained by setting $u_k=const$, $v_k=const,  \ k\ne i, j$, are of Monge-Amp\`ere type, then the full multi-dimensional system (\ref{m00})  must be of  Monge-Amp\`ere type as well. This result can be reformulated geometrically as follows. Let  $X$ be a codimension two submanifold in  ${\bf Gr}(d, V)$. Suppose that the intersection of $X$ with every ${\bf Gr}(3, 5)\subset {\bf Gr}(d, V)$ is a linear section of ${\bf Gr}(3, 5)$. Then $X$ itself must be a linear section.

\section{Concluding remarks}

In this paper we have classified two-component systems of Monge-Amp\`ere type and established  their integrability in all spacial dimensions. It would be interesting to generalise these results to the multi-component case. Let  ${\bf u}=(u^1({\bf x}), \dots, u^n({\bf x}))$, $n\geq 3$,  be functions of $d$ independent variables ${\bf x}=(x^1, \dots, x^d)$. Consider a first-order Monge-Amp\`ere system 
\begin{equation*}
 F^1({\bf u}_1, \dots,  {\bf u}_d)=0, \ \dots, \  F^n({\bf u}_1, \dots,  {\bf u}_d)=0
\end{equation*}
where each $F^i$ is a linear combination of minors of the $n\times d$ Jacobian matrix of  ${\bf u}({\bf x})$.
Geometrically, such systems correspond to sections of  ${\bf Gr}(d, V^{n+d})$ by linear spaces of codimension $n$.
Based on the present paper and the results of \cite{DF, FHK} we can formulate the following conjectures.

\begin{itemize}

\item For $d=3$, the integrability of a Monge-Amp\`ere system is equivalent to its linearisability  (which is equivalent to the property that the corresponding linear space of codimension $n$ is tangential to ${\bf Gr}(3, V^{n+3})$).

\item For $d\geq 4$, the integrability of a Monge-Amp\`ere system (for $n\geq 3$ it will no longer be automatic) is equivalent to the property that the corresponding linear space of codimension $n$ is tangential to ${\bf Gr}(d, V^{n+d})$
along a submanifold which meets every ${\bf Gr}(3, V^{n+3})\subset {\bf Gr}(d, V^{n+d})$.

\end{itemize}
We hope to return to these questions elsewhere.

\section*{Acknowledgements}
We thank  the LMS for their support of BD to Loughborough making this collaboration possible.

\end{document}